\PassOptionsToPackage{unicode}{hyperref}
\PassOptionsToPackage{hyphens}{url}
\PassOptionsToPackage{dvipsnames,svgnames,x11names}{xcolor}
\documentclass[
  12pt]{article}
\newtheorem{assumption}{Assumption}
\usepackage{threeparttable}
\usepackage{multirow} 
\usepackage{tikz}
\usepackage{tikz-cd}
\usepackage{amsmath,amssymb}
\usepackage{iftex}
\ifPDFTeX
  \usepackage[T1]{fontenc}
  \usepackage[utf8]{inputenc}
  \usepackage{textcomp} 
\else 
  \usepackage{unicode-math}
  \defaultfontfeatures{Scale=MatchLowercase}
  \defaultfontfeatures[\rmfamily]{Ligatures=TeX,Scale=1}
\fi
\usepackage{lmodern}
\ifPDFTeX\else  
\fi
\IfFileExists{upquote.sty}{\usepackage{upquote}}{}
\IfFileExists{microtype.sty}{
  \usepackage[]{microtype}
  \UseMicrotypeSet[protrusion]{basicmath} 
}{}
\makeatletter
\@ifundefined{KOMAClassName}{
  \IfFileExists{parskip.sty}{%
    \usepackage{parskip}
  }{
    \setlength{\parindent}{0pt}
    \setlength{\parskip}{6pt plus 2pt minus 1pt}}
}{
  \KOMAoptions{parskip=half}}
\makeatother
\usepackage{xcolor}
\makeatletter
\ifx\paragraph\undefined\else
  \let\oldparagraph\paragraph
  \renewcommand{\paragraph}{
    \@ifstar
      \xxxParagraphStar
      \xxxParagraphNoStar
  }
  \newcommand{\xxxParagraphStar}[1]{\oldparagraph*{#1}\mbox{}}
  \newcommand{\xxxParagraphNoStar}[1]{\oldparagraph{#1}\mbox{}}
\fi
\ifx\subparagraph\undefined\else
  \let\oldsubparagraph\subparagraph
  \renewcommand{\subparagraph}{
    \@ifstar
      \xxxSubParagraphStar
      \xxxSubParagraphNoStar
  }
  \newcommand{\xxxSubParagraphStar}[1]{\oldsubparagraph*{#1}\mbox{}}
  \newcommand{\xxxSubParagraphNoStar}[1]{\oldsubparagraph{#1}\mbox{}}
\fi
\makeatother

\usepackage{longtable,booktabs,array}
\usepackage{calc} 
\usepackage{etoolbox}
\makeatletter
\patchcmd\longtable{\par}{\if@noskipsec\mbox{}\fi\par}{}{}
\makeatother
\IfFileExists{footnotehyper.sty}{\usepackage{footnotehyper}}{\usepackage{footnote}}
\makesavenoteenv{longtable}
\usepackage{graphicx}
\makeatletter
\def\maxwidth{\ifdim\Gin@nat@width>\linewidth\linewidth\else\Gin@nat@width\fi}
\def\maxheight{\ifdim\Gin@nat@height>\textheight\textheight\else\Gin@nat@height\fi}
\makeatother
\setkeys{Gin}{width=\maxwidth,height=\maxheight,keepaspectratio}
\makeatletter
\def\fps@figure{htbp}
\makeatother

\makeatletter
\@ifpackageloaded{caption}{}{\usepackage{caption}}
\AtBeginDocument{%
\ifdefined\contentsname
  \renewcommand*\contentsname{Table of contents}
\else
  \newcommand\contentsname{Table of contents}
\fi
\ifdefined\listfigurename
  \renewcommand*\listfigurename{List of Figures}
\else
  \newcommand\listfigurename{List of Figures}
\fi
\ifdefined\listtablename
  \renewcommand*\listtablename{List of Tables}
\else
  \newcommand\listtablename{List of Tables}
\fi
\ifdefined\figurename
  \renewcommand*\figurename{Figure}
\else
  \newcommand\figurename{Figure}
\fi
\ifdefined\tablename
  \renewcommand*\tablename{Table}
\else
  \newcommand\tablename{Table}
\fi
}
\@ifpackageloaded{float}{}{\usepackage{float}}
\floatstyle{ruled}
\@ifundefined{c@chapter}{\newfloat{codelisting}{h}{lop}}{\newfloat{codelisting}{h}{lop}[chapter]}
\floatname{codelisting}{Listing}

\makeatother
\makeatletter
\@ifpackageloaded{caption}{}{\usepackage{caption}}
\@ifpackageloaded{subcaption}{}{\usepackage{subcaption}}
\makeatother

\ifLuaTeX
  \usepackage{selnolig}  
\fi
\usepackage[]{natbib}
\usepackage{bookmark}

\usepackage{tikz} 
\usepackage{subcaption}
\usepackage{tikz-cd}  \usetikzlibrary{positioning,arrows.meta,fit,shapes.geometric}

\IfFileExists{xurl.sty}{\usepackage{xurl}}{} 
\hypersetup{
  pdftitle={Causal Inference and Evidence Synthesis for Semi-continuous Exposures with Multiple Outcomes},
  pdfauthor={Xiaoya Wang; Richard J. Cook; Yeying Zhu; Tugba Akkaya-Hocagil; R. Colin Carter; Sandra W. Jacobson; Joseph L. Jacobson; Nancy Day; Louise M. Ryan},
  pdfkeywords={Causal effect; Childhood cognition; Estimating equations; Homogeneity tests; Prenatal alcohol exposure; Propensity scores},
  colorlinks=true,
  linkcolor={blue},
  filecolor={Maroon},
  citecolor={Blue},
  urlcolor={Blue},
  pdfcreator={LaTeX via pandoc}}

\newcommand{\anon}{1}

\date{}

\begin{document}

\def\spacingset#1{\renewcommand{\baselinestretch}%
{#1}\small\normalsize} \spacingset{1}

\if1\anon
{
  \title{\bf Causal Inference and Evidence Synthesis for Semi-continuous Exposures with Multiple Outcomes}
  \author{%
  Xiaoya Wang$^{1}$, Richard J. Cook$^{1}$, Yeying Zhu$^{1}$\\
  Tugba Akkaya-Hocagil$^{2}$, R. Colin Carter$^{3}$, Sandra W. Jacobson$^{4}$\\
  Joseph L. Jacobson$^{4}$, Nancy Day$^{5}$, Louise M. Ryan$^{6}$\\[0.7em]
  {\small $^{1}$Department of Statistics and Actuarial Science, University of Waterloo}\\
  {\small $^{2}$Department of Biostatistics, School of Medicine, Ankara University}\\
  {\small $^{3}$Departments of Emergency Medicine and Pediatrics, Columbia University}\\
  {\small $^{4}$Department of Psychiatry and Behavioral Neurosciences, Wayne State University}\\
  {\small $^{5}$Department of Psychiatry, University of Pittsburgh School of Medicine}\\
  {\small $^{6}$School of Mathematical and Physical Sciences, University of Technology Sydney}}
  \maketitle
} \fi

\bigskip
\begin{abstract}
Semi-continuous exposure variables, featuring a point mass at zero and a continuous dose in exposed individuals, arise frequently in epidemiology. In such settings, investigators may wish to distinguish the causal effect of exposure at a fixed dose with no exposure, from the dose-response effect among exposed individuals. We develop a two-stage framework for causal inference with semi-continuous exposures and multiple outcomes. The proposed approach yields outcome-specific estimators for these two causal components, establishes their large sample properties, and provides Wald-type tests for homogeneity across outcomes. We further extend the framework to the multi-cohort setting through hierarchical homogeneity testing and evidence synthesis, enabling principled pooling across cohorts and outcomes. We apply the proposed methods to several cohort studies designed to evaluate the effects of prenatal alcohol exposure on executive function, mathematics achievement, and learning and memory. Simulation studies show that the proposed estimators and testing procedures perform well. The framework provides a principled basis for estimation, comparison, and synthesis of causal effects with multiple outcomes and semi-continuous exposures.
\end{abstract}

\noindent%
{\it Keywords: Causal effect; Childhood cognition; Estimating equations; Homogeneity tests; Prenatal alcohol exposure; Propensity scores.} 
\vfill

\newpage
\spacingset{1.2} 

\section{Introduction}

Propensity scores were originally developed as balancing scores enabling estimation of causal effects with binary exposures \citep{rubin1974random,rosenbaum1983central,rosenbaum1984reducing}. Generalized propensity scores extend this approach to continuous exposures \citep{hirano2004propensity,imai2004causal}. In many areas of public health, however, the exposure distribution has a point mass at zero and a continuous distribution of positive values. In studies of prenatal alcohol exposure (PAE), for example, a substantial proportion of mothers do not drink alcohol during pregnancy, but the level of consumption varies among those who do drink. Semi-continuous exposures such as this raise two distinct questions: status incurs an effect on the outcome, and does the dose of exposure have an effect on the outcome among the exposed.

\cite{akkaya2021propensity} developed a two-part propensity score model in which one component models exposure status and the other models the exposure level among exposed individuals. This was used for combining data across multiple outcomes and studies in a multi-stage meta-analysis in \cite{akkaya2022hierarchical} to estimate the extent of heterogeneity of effects across outcomes and studies; see also \cite{dang2023bayesian} for a Bayesian approach. \cite{li2024use} considered a two-stage approach to analyze which separately examines the effects of the continuous level of exposure among the exposed and the binary exposure status; \cite{wang2025two} cast this approach in the potential outcomes framework and established large sample properties of estimators.  

Our research is motivated by an interdisciplinary collaboration on the effects of PAE on multiple domains of child cognition and behavior \citep{carter2016fetal,jacobson2004maternal,jacobson2008impaired,mattson2019fetal}. In an individual-participant meta-analysis based on prospective U.S. cohorts, adverse effects of PAE were reported across cognitive domains, including executive function, learning and memory, reading, mathematics, and behavior \citep{jacobson2021effects}. This raises two questions: do the causal effects of PAE vary across cognitive domains, and are they homogeneous across cohorts? Separate models for each outcome and cohort cannot answer these, since cognitive outcomes are correlated within a child and pooling across cohorts is interpretable only when the component effects are homogeneous.

For correlated outcomes, multivariate linear mixed models allow flexible modeling of mean and correlation structures \citep{sammel1999multivariate}, and scaled linear mixed models provide a way to estimate both outcome-specific and global effects while accounting for within-subject correlation \citep{lin2000scaled,roy2003scaled}. Bayesian methods accommmodating nested outcomes have been developed \citep{thurston2009bayesian}. More recent semiparametric doubly robust approaches permit causal inference for multiple outcomes of different types or scales while addressing multiplicity \citep{kennedy2019estimating,du2025causal}. Despite these advances, methods for causal inference with a semi-continuous exposure and multiple correlated outcomes remain underdeveloped. In particular, there is little methodology for jointly estimating the effect of exposure status and the dose-response effect among exposed individuals across outcomes, while also formally assessing whether these effects may reasonably be viewed as common within cohort, across cohorts, or both.

We develop a framework for causal inference for a semi-continuous exposure including homogeneity tests to guide evidence synthesis across outcomes and cohorts. We first consider the setting of multiple outcomes in a single cohort and use generalized estimating equations \citep{liang1986longitudinal} to obtain outcome-specific estimators for the dose-response effect among exposed individuals and the effect of exposure at a reference dose compared to no exposure. We then address the multi-cohort setting, where interest lies in assessing heterogeneity and evidence synthesis across studies. When effects appear homogeneous, pooling yields precise estimates and interpretable summaries; when heterogeneity is evident, it can reveal which cohorts or domains are differentially affected by exposure.

The remainder of the article is organized as follows. In Section~\ref{sec-2}, we introduce the motivating study of prenatal alcohol exposure and childhood cognition and describe the data from the several cohorts. In Section~\ref{sec-3} we define the causal estimands for a semi-continuous exposure with multiple outcomes. Estimation procedures are then described along with the large sample properties of estimators; homogeneity tests are also presented. The multi-cohort setting and corresponding hierarchical homogeneity testing and evidence-synthesis procedures are addressed in Section \ref{sec-4}. The results of simulation studies assessing the finite sample performance of the proposed methods are given in \textit{Supplementary Material}. In Section~\ref{sec-5}, we apply the methods first to a single-cohort study and then illustrate the synthesis across five cohorts. Section~\ref{sec-7} concludes with a discussion.

\section{Prenatal Alcohol Exposure and Childhood Cognition}
\label{sec-2}
\begin{figure}[!t]
\centering
\includegraphics[]{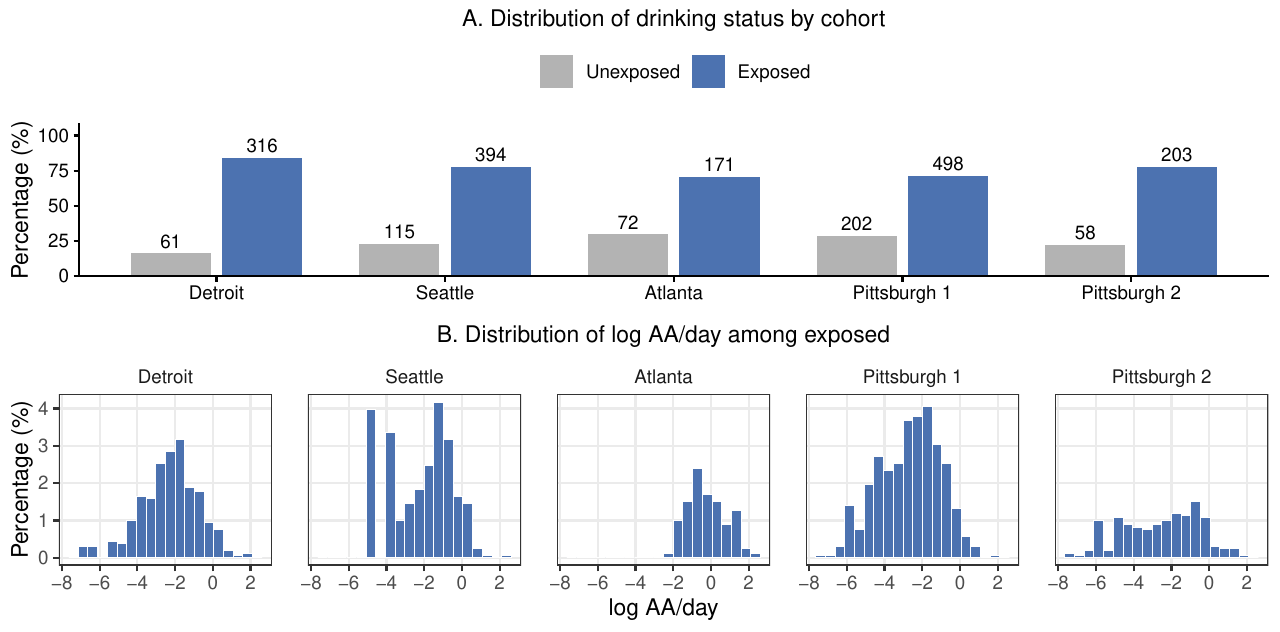}
\caption{Cohort-specific distributions of prenatal alcohol exposure in the motivating study.}
\label{fig-1}
\end{figure} 

\begin{table}[t]
\centering
\footnotesize
\caption{Cohort-specific summaries of prenatal alcohol exposure.}
\label{tab:cohort_pae_summary}
\resizebox{\columnwidth}{!}{%
\begin{tabular}{lcccc}
\hline
Cohort & $N$ & PAE $>$ 0, $n$ (\%) & Mean log(AA/day) among exposed & Median log(AA/day) among exposed (IQR) \\
\hline
Detroit      & 377 & 316 (83.8) & -2.307 & -2.207 (-3.277, -1.352) \\
Seattle      & 509 & 394 (77.4) & -2.174 & -1.833 (-3.912, -0.923) \\
Atlanta    & 243 & 171 (70.4) & -0.234 & -0.154 (-0.847, 0.452) \\
Pittsburgh-1 & 700 & 498 (71.1) & -2.687 & -2.528 (-3.912, -1.497) \\
Pittsburgh-2 & 261 & 203 (77.8) & -2.557 & -2.270 (-4.194, -0.876) \\
\hline 
Overall      & 2090 & 1582 (75.7) & -2.202 & -1.970 (-3.540, -0.886) \\
\hline
\end{tabular}}
\end{table}

We consider data from several NIH-funded cohort studies designed to evaluate the impact of maternal alcohol use during pregnancy on child cognitive and behavioral development \citep{jacobson2021effects}. Here, we consider the five cohorts with the broadest coverage of the cognitive domains of interest: one cohort from Detroit \citep{jacobson1993teratogenic}, one from Seattle \citep{streissguth1981seattle}, one from Atlanta \citep{coles1997comparison}, and two from Pittsburgh \citep{day1991prenatal,richardson1999growth}; over 2,090 mother-child pairs are represented. Additional cohort-specific summaries of baseline covariates and the extent of missing data are provided in Section 1 of the \textit{Supplementary Material}.

Prenatal alcohol exposure is recorded as the average ounces of absolute alcohol (AA) consumed per day over the pregnancy \citep{jacobson2021effects}. Many mothers report no alcohol use during pregnancy, resulting in a semi-continuous exposure. If $T$ is the average ounces of absolute alcohol consumed per day, we let $A = I(T > 0)$ indicate any PAE and $D = \log T$ denote dose (i.e., $\log$ AA/day) defined for those exposed. Table~\ref{tab:cohort_pae_summary} provides cohort-specific numerical summaries of PAE, including the proportion exposed and the distribution of dose among those exposed, while Figure~\ref{fig-1} displays the corresponding cohort-specific exposure distributions. 

This application is motivated by four scientific questions: 1) What is the effect of a reference PAE dose compared with no exposure on child cognition? 2) What is the dose-response effect of PAE on cognitive outcomes in exposed children? 3) Are these effects homogeneous across cognitive domains? 4) For a given cognitive domain, are the causal effects homogeneous across cohorts, thereby justifying evidence synthesis? 

\begin{figure}[!t]
\centering
\includegraphics[width=1\textwidth]{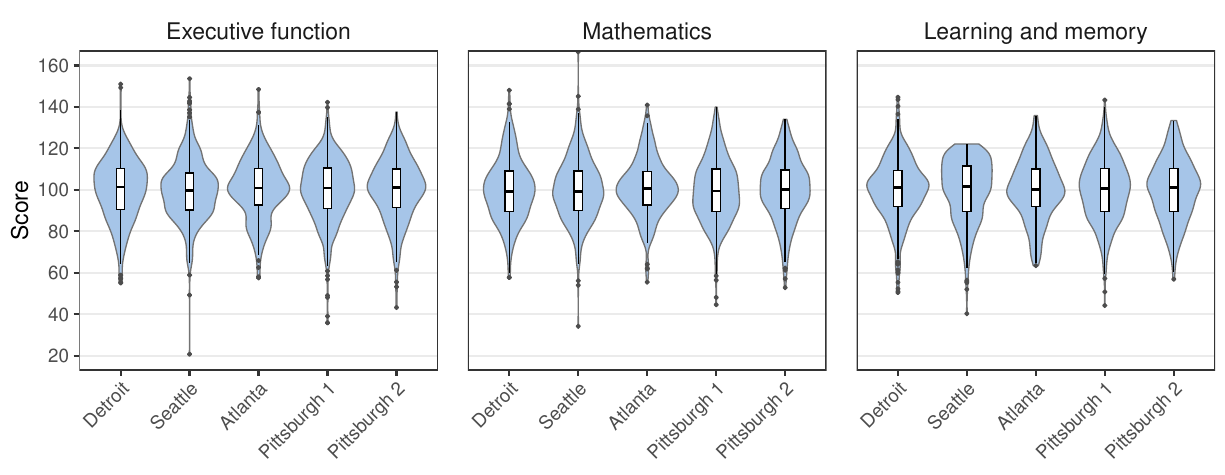}
\caption{Distributions of harmonized cognitive domain scores by cohort in the motivating study.}
\label{fig-2}
\end{figure}

We focus on three domain-level cognitive outcomes: executive function (EF), mathematics achievement (Math), and learning and memory (LM). The EF, Math, and LM outcomes are harmonized latent domain scores constructed from multiple neuropsychological assessments administered across childhood, with higher values indicating better performance; see Figure~\ref{fig:2A} for an illustration of the structural equation model used to generate the domain-level latent outcomes. Figure~\ref{fig-2} displays the cohort-specific marginal distributions of the three harmonized domain scores; they are broadly comparable across cohorts, while still showing some differences in spread and shape. We consider five cohorts which provide harmonized information on the three domains of interest. Across the five cohorts, EF, Math, and LM were obtained for 2,029 (97.1\%), 2,051 (98.1\%), and 1,962 (93.9\%) children, respectively, and 1,924 (92.1\%) had all three domain scores observed. Missing baseline covariates and incomplete domain scores were handled by multiple imputation within cohorts; see Section 3 of the \textit{Supplementary Material}.

Baseline maternal and pregnancy-related covariates include demographic characteristics, reproductive history, smoking and other substance-use measures, and additional cohort-specific maternal or pregnancy factors. The full lists of cohort-specific covariates are provided in Section 3 of the \textit{Supplementary Material}. Since the exact covariate sets differed across cohorts, propensity scores are developed within each cohort \citep{jacobson2021effects}; 

\begin{figure}[t]
\centering

\begin{subfigure}[t]{0.48\textwidth}
\centering 
\begin{tikzpicture}[
  baseline=(current bounding box.north),
    inner sep=0pt, outer sep=0pt,  
  node distance=12mm and 14mm,
  latent/.style={ellipse, draw, align=center, inner sep=2pt},
  observed/.style={rectangle, draw, minimum width=15mm, minimum height=6mm, align=center},
  group/.style={draw, dashed, rounded corners, inner sep=4pt},
  arrow/.style={-Latex}
]
\node[latent] (Cog) {Cognition};

\node[latent, below left=10mm and 2mm of Cog] (EF) {EF};
\node[latent, below=8.5mm of Cog]              (Math) {Math};
\node[latent, below right=10mm and 2mm of Cog](LM) {LM};

\node[observed,  below left=10mm and 2mm of EF] (EFbox) {Tests 1--7};
\node[observed,   below=9mm of Math]           (Mbox)  {Tests 1, 2};
\node[observed,   below right=10mm and 2mm of LM](LMbox) {Tests 1--6};

\draw[arrow] (Cog) -- (EF);
\draw[arrow] (Cog) -- (Math);
\draw[arrow] (Cog) -- (LM);

\draw[arrow] (EF)   -- (EFbox.north);
\draw[arrow] (Math) -- (Mbox.north);
\draw[arrow] (LM)   -- (LMbox.north);

\tikzset{covar/.style={<->, >=Latex, looseness=8, shorten <=1.5pt, shorten >=1.5pt}}
\draw[covar] (Cog)  to[out=60,  in=120] (Cog);
\draw[covar] (EF)   to[out=155, in=110] (EF);
\draw[covar] (Math) to[out=145, in=110] (Math);
\draw[covar] (LM)   to[out=30,  in=70]  (LM);
\draw[covar] (EFbox) to[out=300, in=240] (EFbox);
\draw[covar] (Mbox)  to[out=300, in=240] (Mbox);
\draw[covar] (LMbox) to[out=300, in=240] (LMbox);
\end{tikzpicture}
\caption{\scriptsize SEM for deriving domain-level outcomes in the Detroit cohort. Rectangles are observed test groups; ellipses, latent variables; single-headed arrows, factor loadings; double-headed arrows between test boxes, residual correlations; self-loops, factor variances.}
\label{fig:2A}
\end{subfigure}
\hfill
\begin{subfigure}[t]{0.48\textwidth}
 \centering \vspace*{-10.5ex} \hspace*{-9ex}
\begin{tikzcd}[baseline=(current bounding box.north), 
row sep=-2em, column sep=2em]
\mathbf{X}_1 & T\\
\mathbf{X}_3 && \left(\begin{matrix} Y_1\\ Y_2\\ \vdots\\ Y_J \end{matrix}\right) \\
\mathbf{X}_2 & A
\arrow[from=1-1, to=1-2]
\arrow[from=1-2, to=2-3]
\arrow[from=3-1, to=3-2]
\arrow[from=3-2, to=2-3]
\arrow[from=2-1, to=3-2]
\arrow[from=2-1, to=1-2]
\arrow[from=3-2, to=1-2]
\arrow[bend left=60, from=1-1, to=2-3]
\arrow[bend right=60, from=3-1, to=2-3]
\arrow[out=130, in=90, looseness=1.7, from=2-1, to=2-3]
\end{tikzcd}
\caption{\scriptsize  Directed acyclic graph (DAG) for a semi-continuous exposure.}
\label{fig:2B}
\end{subfigure}

 \caption{Overview of the problem: Panel A shows the SEM measurement model used to construct domain-level outcomes; panel B shows the causal DAG for the exposure.}
\label{fig:overview}
\end{figure}

\section{Methods for Two-stage Estimation}
\label{sec-3}
Here we describe a two-stage estimation procedure. We use propensity score regression adjustment in Stage I to mitigate confounding when estimating the causal dose-response effect among the exposed. In Stage II, the effect of the exposure at a reference dose compared to no exposure can be estimated via propensity score regression \citep{vansteelandt2014regression}, inverse probability weighting (IPW) \citep{robins2000marginal}, or augmented inverse probability weighting (AIPW) \citep{bang2005doubly}. We focus on AIPW due to its doubly robust property in Stage II, but details on alternative Stage II estimators are given in Section 2 of the \textit{Supplementary Material}. 

\subsection{Semi-continuous exposure setup and causal estimands}
\label{sec-3.1}

Let $i$ label individuals in a sample of $n$ independent individuals, $i=1,\ldots,n$, and $\mathbf{Y}_i=(Y_{i1},\ldots,Y_{iJ})'$ denote a $J\times 1$ vector of continuous outcomes. Let $A_i=1$ if a woman consumes alcohol during pregnancy and $A_i=0$ otherwise; we refer to $A_i$ as the exposure status. Let $T_i$ denote the non-negative daily volume of alcohol consumed when $A_i=1$ and let $D_i=\log T_i$ be the ``dose'' of exposure if $A_i=1$; $D_i$ is undefined if $A_i=0$.

The confounders for the dose-response analyses among exposed individuals are $\mathbf{Z}_{i1}$, and $\mathbf{Z}_{i2}$ denotes the confounders for the effect of exposure status. To accommodate overlap in these two covariate sets we write $\mathbf{X}_i=(\mathbf{X}_{i1}',\mathbf{X}_{i2}',\mathbf{X}_{i3}')'$, where $\mathbf{X}_{i1}$ are covariates associated with the continuous exposure level and the outcomes among exposed individuals, $\mathbf{X}_{i2}$ are covariates associated with exposure status and the outcomes, and $\mathbf{X}_{i3}$ contains covariates associated with both components of the exposure and outcomes. We then define $\mathbf{Z}_{i1}=(\mathbf{X}_{i1}',\mathbf{X}_{i3}')'$ and $\mathbf{Z}_{i2}=(\mathbf{X}_{i2}',\mathbf{X}_{i3}')'$. The directed acyclic graph in Figure~\ref{fig:2B} summarizes the relationships among exposure, dose, covariates, and outcomes. Let $\mathcal{D}_i=(\mathbf{Y}_i,\mathbf{X}_i,A_i,D_i)$ denote the observed data for each individual $i$, and let $\mathcal{D}=\{\mathcal{D}_i:i=1,\ldots,n\}$ denote the full sample.

To formalize the causal effects of interest, let $\mathbf{Y}_i(a,d)=\{Y_{i1}(a,d),\ldots,Y_{iJ}(a,d)\}'$ denote the vector of potential outcomes under exposure status $a$ and dose $d$. For exposed individuals, $\mathbf{Y}_i(1,d)$ denotes the potential outcome vector if the individual is exposed at dose $d$. For unexposed individuals, the dose is undefined, but for notational convenience we write the corresponding potential outcome vector as $\mathbf{Y}_i(0,0)$.

We consider two distinct causal effects. First, for a given dose level $d$, we define the vector of average causal dose-response effects among exposed individuals as
\begin{equation}
\label{eq:dose_response_vector}
E\{\mathbf{Y}_i(1,d+1)\mid A_i=1\}
-
E\{\mathbf{Y}_i(1,d)\mid A_i=1\}.
\end{equation}
This estimand describes how the mean of outcomes changes with a one-unit increase in dose among exposed individuals.

Second, to define the causal effect of exposure status, we specify a reference dose $c>0$. The vector of average causal effects is then 
\begin{equation}
\label{eq:status_effect_vector}
E\{\mathbf{Y}_i(1,c)\}
-
E\{\mathbf{Y}_i(0,0)\}.
\end{equation}
We refer to \eqref{eq:status_effect_vector} as the causal effect of exposure status at reference dose $c$ compared to no exposure. In practice $c$ may be set to a clinically meaningful exposure level; setting $c=\log 1 =0$ corresponds to comparing exposure at 1 ounce of absolute alcohol per day on the original scale, with that of the unexposed.

Identification of these causal effects relies on the usual assumptions of consistency, conditional exchangeability, and positivity \citep{rubin1980randomization,rosenbaum1983central,rubin1990application,cole2008constructing}.

\begin{assumption}[Consistency and no interference]
\label{ass:sutva_multi}
There is no interference between individuals; and $\mathbf{Y}_i=\mathbf{Y}_i(0,0)$ if $A_i=0$,
$\mathbf{Y}_i=\mathbf{Y}_i(1,d)$ if $A_i=1$ and $D_i=d$.
\end{assumption}

\begin{assumption}[Conditional exchangeability]
\label{ass:exchange_multi}
For all $d$ in the dose range of interest, $\mathbf{Y}_i(1,d)\perp D_i \mid (A_i=1,\mathbf{Z}_{i1})$ and 
$\{\mathbf{Y}_i(0,0),\mathbf{Y}_i(1,d)\}\perp A_i \mid \mathbf{Z}_{i2}$.
\end{assumption}

\begin{assumption}[Positivity]
\label{ass:positivity_multi}
For covariate values with positive density, $0<P(A_i=1\mid \mathbf{Z}_{i2})<1$ and $f(D_i\mid A_i=1,\mathbf{Z}_{i1})>0$
for all $d$ in the dose range of interest.
\end{assumption}

Consistency links the observed outcome vector to the potential outcome vector under the observed exposure and rules out interference between individuals. Conditional exchangeability states that, after adjustment for measured baseline covariates, exposure status and dose among the exposed are independent of the corresponding potential outcomes. Positivity requires that each individual has a positive probability of either exposure status and, among the exposed, a positive conditional density over the dose range of interest. Under these assumptions, the causal effects in \eqref{eq:dose_response_vector} and \eqref{eq:status_effect_vector} are identifiable.

\subsection{Model Setup}
\label{sec-3.2}
In the data generation model, we assume that the response vector $\mathbf{Y}_i$ follows a multivariate linear regression model with a semi-continuous exposure:
{\footnotesize
\begin{align}
\label{equa-DGmodel}
    \left(\begin{matrix}
        Y_{i1}\\
        Y_{i2}\\
        \vdots\\
        Y_{iJ}
    \end{matrix}\right)
    &=
    \left(\begin{matrix}
        \theta_{01}\\
        \theta_{02}\\
        \vdots\\
        \theta_{0J}
    \end{matrix}\right)
    +
    \left(\begin{matrix}
        \theta_{11}\\
        \theta_{12}\\
        \vdots\\
        \theta_{1J}
    \end{matrix}\right) A_i(D_i-c)
    +
    \left(\begin{matrix}
        \theta_{21}\\
        \theta_{22}\\
        \vdots\\
        \theta_{2J}
    \end{matrix}\right) A_i
    +
    \left(\begin{matrix}
        \boldsymbol{\theta}_{311}' & \boldsymbol{\theta}_{321}' & \boldsymbol{\theta}_{331}'\\
        \boldsymbol{\theta}_{312}' & \boldsymbol{\theta}_{322}' & \boldsymbol{\theta}_{332}'\\
        \vdots & \vdots & \vdots\\
        \boldsymbol{\theta}_{31J}' & \boldsymbol{\theta}_{32J}' & \boldsymbol{\theta}_{33J}'
    \end{matrix}\right)
    \left(\begin{matrix}
        \mathbf{X}_{i1}\\
        \mathbf{X}_{i2}\\
        \mathbf{X}_{i3}
    \end{matrix}\right)
    +
    \left(\begin{matrix}
        E_{i1}\\
        E_{i2}\\
        \vdots\\
        E_{iJ}
    \end{matrix}\right).
\end{align}
}
Here, $\mathbf{E}_i=(E_{i1},E_{i2},\ldots,E_{iJ})'$ is a $J$-dimensional error term assumed to follow a multivariate distribution with mean vector $\mathbf{0}$ and covariance matrix $\boldsymbol{\Sigma}$, which accommodates correlation among the $J$ responses. We write $\boldsymbol{\theta}_0=(\theta_{01},\ldots,\theta_{0J})'$, $\boldsymbol{\theta}_1=(\theta_{11},\ldots,\theta_{1J})'$, and $\boldsymbol{\theta}_2=(\theta_{21},\ldots,\theta_{2J})'$. We further let $\boldsymbol{\theta}_{3}=(\boldsymbol{\theta}_{31},\boldsymbol{\theta}_{32},\boldsymbol{\theta}_{33})$, where $\boldsymbol{\theta}_{3r}=(\boldsymbol{\theta}_{3r1},\ldots,\boldsymbol{\theta}_{3rJ})'$ contains the coefficients corresponding to $\mathbf{X}_{ir}$ for $r=1,2,3$ and $j=1,\ldots,J$. Under Assumptions~\ref{ass:sutva_multi}--\ref{ass:positivity_multi}, if the data generation model in \eqref{equa-DGmodel} is correctly specified, then $\boldsymbol{\theta}_1$ and $\boldsymbol{\theta}_2$ are of primary interest, as they characterize the causal dose-response effect among the exposed and the causal effect of exposure at the reference dose compared with no exposure across outcomes.

To address the semi-continuous nature of the exposure variable, we adopt a linear model for the dose $D_i$ among exposed individuals and a logistic regression model for the binary exposure status $A_i$. Let 
\begin{align}
\label{equa-contexpo}
S_{ 1}(\mathbf{Z}_{i1};\boldsymbol{\alpha}_1)
=
E(D_i\mid A_i=1,\mathbf{Z}_{i1};\boldsymbol{\alpha}_1)
=
\bar{\mathbf{Z}}_{i1}'\boldsymbol{\alpha}_1,
\end{align}
where $\bar{\mathbf{Z}}_{i1}=(1,\mathbf{Z}_{i1}')'=(1,\mathbf{X}_{i1}',\mathbf{X}_{i3}')'$ and $\boldsymbol{\alpha}_1=(\alpha_{10},\boldsymbol{\alpha}_{11}',\boldsymbol{\alpha}_{12}')'$, with $\boldsymbol{\alpha}_{11}=(\alpha_{111},\ldots,\alpha_{11k_1})'$ and $\boldsymbol{\alpha}_{12}=(\alpha_{121},\ldots,\alpha_{12k_3})'$. For the data generating process, let $W_i$ denote an error term with $W_i\perp \mathbf{Z}_{i1}|A_i=1$ and $W_i\sim N(0,\sigma_W^2)$, so $D_i=S_{ 1}(\mathbf{Z}_{i1};\boldsymbol{\alpha}_1)+W_i$.

For exposure status, we let 
\begin{align}
\label{equa-binaryexpo}
\log\left\{\frac{P(A_i=1\mid \mathbf{Z}_{i2};\boldsymbol{\alpha}_2)}{1-P(A_i=1\mid \mathbf{Z}_{i2};\boldsymbol{\alpha}_2)}\right\}
=
\bar{\mathbf{Z}}_{i2}'\boldsymbol{\alpha}_2
\end{align}
where $\bar{\mathbf{Z}}_{i2}=(1,\mathbf{Z}_{i2}')'=(1,\mathbf{X}_{i2}',\mathbf{X}_{i3}')'$ and $\boldsymbol{\alpha}_2=(\alpha_{20},\boldsymbol{\alpha}_{21}',\boldsymbol{\alpha}_{22}')'$, with $\boldsymbol{\alpha}_{21}=(\alpha_{211},\ldots,\alpha_{21k_2})'$ and $\boldsymbol{\alpha}_{22}=(\alpha_{221},\ldots,\alpha_{22k_3})'$. We then set $S_2 (\mathbf{Z}_{i2};\boldsymbol{\alpha}_2)=P(A_i=1\mid \mathbf{Z}_{i2};\boldsymbol{\alpha}_2)$ as the corresponding propensity score for exposure status.

\subsection{Two-stage estimation}
\label{sec-3.3}
Here we extend the approach of \cite{wang2025two} to accommodate multiple outcomes. Stage~I targets the causal dose-response effect among exposed individuals, and Stage~II targets the causal effect of exposure status at a reference dose. In both stages, we use generalized estimating equations (GEE) \citep{liang1986longitudinal} with a working independence correlation structure. The robust sandwich covariance estimate ensures asymptotically valid inference \citep{liang1986longitudinal,zeger1986longitudinal}.

\subsubsection{Stage I: Estimation of Causal Dose-response Effects among Exposed}
In Stage~I, we estimate the causal dose-response effect among exposed individuals using propensity score regression adjustment. This stage combines a model for dose among the exposed with an outcome regression that adjusts for the fitted dose score.
We specify the regression adjustment model with the propensity score
{\small\begin{equation}
\label{equa-3.1}
E\left\{\left.\left(\begin{matrix}
    Y_{i1}\\
    Y_{i2}\\
    \vdots\\
    Y_{iJ}
\end{matrix}\right)\right| A_i = 1, D_i, S_{1}(\boldsymbol{Z}_{i1}; \boldsymbol{\alpha}_1)\right\}= \left(\begin{matrix}
    \gamma_{101}\\
    \gamma_{102}\\
    \vdots\\
    \gamma_{10J}
\end{matrix}\right) + \left(\begin{matrix}
    \gamma_{111}\\
    \gamma_{112}\\
    \vdots\\
    \gamma_{11J}
\end{matrix}\right)  D_i +\left(\begin{matrix}
    \gamma_{121}\\
    \gamma_{122}\\
    \vdots\\
    \gamma_{12J}
\end{matrix}\right)   S_{1}(\boldsymbol{Z}_{i1}; \boldsymbol{\alpha}_1).
\end{equation}}
Let $\boldsymbol{\gamma}_1=(\boldsymbol{\gamma}_{10}',\boldsymbol{\gamma}_{11}',\boldsymbol{\gamma}_{12}')'$, where $\boldsymbol{\gamma}_{10}=(\gamma_{101},\ldots,\gamma_{10J})'$, $\boldsymbol{\gamma}_{11}=(\gamma_{111},\ldots,\gamma_{11J})'$, and $\boldsymbol{\gamma}_{12}=(\gamma_{121},\ldots,\gamma_{12J})'$, and let $\boldsymbol{\phi}_1=(\boldsymbol{\gamma}_1',\boldsymbol{\alpha}_1')'$. The Stage~I generalized estimating equation \citep{liang1986longitudinal} for $\boldsymbol{\gamma}_1$ is
\begin{equation}
\label{newequa-3.3}
    \mathbf{U}_{11}(\mathcal{D};\boldsymbol{\phi}_1)=\sum_{i=1}^{n}\mathbf{U}_{i11}(\mathcal{D}_i;\boldsymbol{\phi}_1)=\sum_{i=1}^{n}A_i\mathbf{H}_{i11}(\boldsymbol{\phi}_1)\mathbf{I}\left\{\mathbf{Y}_i-\boldsymbol{\mu}_{i11}(\boldsymbol{\phi}_1)\right\}=\mathbf{0}
\end{equation}
 under a working independence correlation matrix where $\mathbf{H}_{i11}(\boldsymbol{\phi}_1)={\partial \boldsymbol{\mu}_{i11}(\boldsymbol{\phi}_1)}/{\partial\boldsymbol{\gamma}_1'}$, $\boldsymbol{\mu}_{i11}(\boldsymbol{\phi}_1)=\boldsymbol{\gamma}_{10} + \boldsymbol{\gamma}_{11} D_i + \boldsymbol{\gamma}_{12} S_1(\boldsymbol{Z}_{i1}; \boldsymbol{\alpha}_1)$ and $\mathbf{I}$ is a $J\times J$ identity matrix. The estimating equation for $\boldsymbol{\alpha}_1$ in the exposure model $D_i| A_i=1,\mathbf{Z}_{i1}$ is
\begin{equation}
\label{newequa-3.4}
\mathbf{U}_{12}(\mathcal{D};\boldsymbol{\alpha}_1)=\sum^n_{i=1}\mathbf{U}_{i12}(\mathcal{D}_i;\boldsymbol{\alpha}_1)=\sum^n_{i=1}A_i\mathbf{H}_{i2}(\boldsymbol{\alpha}_1)\left\{D_i-{\mu}_{i12}(\boldsymbol{\alpha}_1)\right\}=\mathbf{0}  
\end{equation}
with $ \mathbf{H}_{i2}(\boldsymbol{\alpha}_1)={\partial  {\mu}_{i12}(\boldsymbol{\alpha}_1)}/{\partial\boldsymbol{\alpha}_1'}$ and ${\mu}_{i12}(\boldsymbol{\alpha}_1)=\bar{\mathbf{Z}}_{i1}'\boldsymbol{\alpha}_1$. Solving \eqref{newequa-3.3} and \eqref{newequa-3.4} yields $\hat{\boldsymbol{\phi}}_1$, including $\hat{\boldsymbol{\gamma}}_{11}$, the estimated vector of causal dose-response effects among exposed individuals.
 
\subsubsection{Stage II: Estimation of the Effects of Exposure Status}
In Stage~II, we estimate the causal effect of exposure status at the reference dose $c$ compared to unexposed, treating $\hat{\boldsymbol{\gamma}}_{11}A_i(D_i-c)$ as an offset term. Estimation is carried out in two stages, so an estimate of $\boldsymbol{\gamma}_{11}$ will be used $\hat{\boldsymbol{\gamma}}_{11}A_i(D_i-c)$ as a genuine offset (i.e. a number) at Stage II, but to discuss the large sample properties of the resulting estimators, we represent the offset in terms of the unknown $\gamma_{11}$ (e.g., ${\boldsymbol{\gamma}}_{11}A_i(D_i-c)$) in what follows. We use an augmented inverse probability weighted estimating equation 
\begin{equation}
\label{equa-estimatingequationAIPW}
    {{\mathbf{U}}}_{21}(\mathcal{D};\boldsymbol{\gamma}_{1},\boldsymbol{\phi}_{2})=
    \sum^n_{i=1}{{\mathbf{U}}}_{i21}(\mathcal{D}_i;\boldsymbol{\gamma}_{1},\boldsymbol{\phi}_{2})=\mathbf{0}  
\end{equation}
where ${{\mathbf{U}}}_{i21}(\mathcal{D}_i;\boldsymbol{\gamma}_{1},\boldsymbol{\phi}_{2})$ is 
$$
\begin{aligned}
   \sum_{a=0}^1&w_i\left(a ; \boldsymbol{\alpha}_2\right)\mathbf{H}_{i21}(a;\boldsymbol{\gamma}_2)\mathbf{I}\big[\mathbf{Y}_i-\boldsymbol{\mu}_{21}(a;\boldsymbol{\gamma}_2)+\mathbf{O}_{i}(\boldsymbol{\gamma}_{11})\big]-\left\{w_i\left(a ; \boldsymbol{\alpha}_2\right)-1\right\}\boldsymbol{g}_i(a;\boldsymbol{\zeta},\boldsymbol{\gamma}_2),
\end{aligned}
$$ $\mathbf{H}_{i21}(a;\boldsymbol{\gamma}_2)={\partial \boldsymbol{\mu}_{21}(a;\boldsymbol{\gamma}_2)}/{\partial \boldsymbol{\gamma}'_2}$ is a $2J\times J$ matrix, $\mathbf{O}_{i}(\boldsymbol{\gamma}_{11})=\boldsymbol{\gamma}_{11}A_i(D_i-c)$ and $\boldsymbol{\mu}_{21}\left(a ; \boldsymbol{\gamma}_2\right)=\boldsymbol{\gamma}_{20}+\boldsymbol{\gamma}_{21} a$ are $J\times1$ vectors for $a=0,1$. Here $\boldsymbol{\phi}_{2}=(\boldsymbol{{\gamma}}_2',\boldsymbol{{\alpha}}_2',\boldsymbol{{\zeta}}')$ and \begin{align} 
w_i\left(a ; \boldsymbol{\alpha}_2\right)=\frac{I\left(A_i=a\right)}{S_2\left(\mathbf{Z}_{i 2} ; \boldsymbol{\alpha}_2\right)^a\left\{1-S_2\left(\mathbf{Z}_{i 2} ; \boldsymbol{\alpha}_2\right)\right\}^{1-a}} \nonumber
\end{align}
for $a=0,1$ and 
$\boldsymbol{g}_i(a;\boldsymbol{\zeta},\boldsymbol{\gamma}_2)=\mathbf{H}_{i21}(a;\boldsymbol{\gamma}_2)\mathbf{I}\left\{\mathbf{m}_{ia}(\boldsymbol{\zeta})-\boldsymbol{\mu}_{21}(a;\boldsymbol{\gamma}_2)\right\},$
where $\mathbf{m}_{ia}(\boldsymbol{\zeta})$ represents the imputation model for those with $a=1$ or $0$ respectively:
$$\mathbf{m}_{ia}(\boldsymbol{\zeta}_a)=E\left\{\mathbf{Y}_i- {\boldsymbol{\gamma}}_{11}a(D-c)|\mathbf{X}_i;\boldsymbol{\zeta}_a\right\}= \boldsymbol{\zeta}_{a0}+\boldsymbol{\zeta}_{a1}\mathbf{X}_{i1}+\boldsymbol{\zeta}_{a2}\mathbf{X}_{i2}+\boldsymbol{\zeta}_{a3}\mathbf{X}_{i3}$$ with $\boldsymbol{\zeta}_{a0}=(\zeta_{a01},...,\zeta_{a0J})'$ and the $J \times k_r$ coefficient matrices $\boldsymbol{\zeta}_{ar} = (\boldsymbol{\zeta}_{ar1}, \ldots, \boldsymbol{\zeta}_{arJ})'$ for $r=1,2,3$. We let $\boldsymbol{\zeta}_a=\text{vec}(\boldsymbol{\zeta}_{a0}',\boldsymbol{\zeta}_{a1}',\boldsymbol{\zeta}_{a2}',\boldsymbol{\zeta}_{a3}')'$ and $\boldsymbol{\zeta}=(\boldsymbol{\zeta}_1',\boldsymbol{\zeta}_0')'$.

The estimating equation to obtain $\boldsymbol{\alpha}_2$ in the Stage II propensity score model is
\begin{equation}
\label{new.estimating.equation3.3}
\mathbf{U}_{22}\left(\mathcal{D} ; \boldsymbol{\alpha}_2\right)=\sum^n_{i=1}\mathbf{U}_{i22}\left(\mathcal{D}_i; \boldsymbol{\alpha}_2\right)=\mathbf{0}
\end{equation}
where
$$\mathbf{U}_{i22}\left(\mathcal{D}_i; \boldsymbol{\alpha}_2\right)=\frac{\partial S_2(\mathbf{Z}_{i2};\boldsymbol{\alpha}_2)}{\partial\boldsymbol{\alpha}_2}\frac{1}{S_2(\mathbf{Z}_{i2};\boldsymbol{\alpha}_2)\left\{1-S_2(\mathbf{Z}_{i2};\boldsymbol{\alpha}_2)\right\}}\left\{A_i-S_2(\mathbf{Z}_{i2};\boldsymbol{\alpha}_2)\right\}.$$ To obtain $\hat{\boldsymbol{\zeta}}$ we consider this estimating equation
\begin{equation}
\label{equa-imputedmodel}
    \mathbf{U}_{23}(\mathcal{D};\boldsymbol{\gamma}_{1},\boldsymbol{\zeta})=
    \sum^n_{i=1}\mathbf{U}_{i23}(\mathcal{D}_i;\boldsymbol{\gamma}_{1},\boldsymbol{\zeta})
\end{equation}
where $\mathbf{U}_{i23}(\mathcal{D}_i;\boldsymbol{\gamma}_{1},\boldsymbol{\zeta})=\left\{\mathbf{U}_{i231}'(\mathcal{D}_i;\boldsymbol{\gamma}_{1},\boldsymbol{\zeta}_1),\mathbf{U}_{i232}'(\mathcal{D}_i;\boldsymbol{\zeta}_0)\right\}'$
with 
\[
\mathbf{U}_{i 231}\left(\mathcal{D}_i ; \boldsymbol{\gamma}_1, \boldsymbol{\zeta}_1\right)=A_i\mathbf{F}_{i1}(\boldsymbol{\zeta}_{1})\mathbf{I}\left\{\mathbf{Y}_i- {\boldsymbol{\gamma}}_{11}(D_i-c)-\mathbf{m}_{i1}\left(\boldsymbol{\zeta}_1\right)\right\} \]
with $\mathbf{F}_{i1}(\boldsymbol{\zeta}_{1})= {\partial \mathbf{m}_{i1}\left(\boldsymbol{\zeta}_1\right)}/{\partial \boldsymbol{\zeta}_1'}$ and
$$\mathbf{U}_{i 232}\left(\mathcal{D}_i ; \boldsymbol{\zeta}_0\right)=(1-A_i)\mathbf{F}_{i2}(\boldsymbol{\zeta}_0)\mathbf{I}\left\{\mathbf{Y}_i-\mathbf{m}_{i0}(\boldsymbol{\zeta}_0)\right\}$$ with $ \mathbf{F}_{i2}(\boldsymbol{\zeta}_0)={\partial \mathbf{m}_{i0}\left(\boldsymbol{\zeta}_0\right)}/{\partial \boldsymbol{\zeta}_0}'$
specified under a working independence assumption.

In practice, we first fit the Stage~I dose-response model among exposed individuals and obtain $\hat{\gamma}_{11}$, the estimated vector of causal dose-response effects. We then substitute $\hat{\gamma}_{11}A_i(D_i-c)$ as an offset in the Stage~II estimating equations \eqref{equa-estimatingequationAIPW} and solve for the parameters governing the causal effects of exposure status. 
\subsubsection{Joint Estimating Equations and Large Sample Properties}
Combining \eqref{newequa-3.3}, \eqref{newequa-3.4}, \eqref{new.estimating.equation3.3}, \eqref{equa-estimatingequationAIPW}, and \eqref{equa-imputedmodel} yields the joint estimating equation for the two-stage procedure:
\begin{align}
    \label{equa-13}
    \mathbf{U}(\boldsymbol{\Omega})=\sum_{i=1}^{n}\mathbf{U}_i(\boldsymbol{\Omega})=\mathbf{0},
\end{align}
where
\[
\mathbf{U}_i(\boldsymbol{\Omega})
=
\left(
\mathbf{U}_{i1}'\left(\mathcal{D}_i ; \boldsymbol{\phi}_1\right),
\mathbf{U}_{i2}'\left(\mathcal{D}_i ; \boldsymbol{\gamma}_1, \boldsymbol{\phi}_2\right)
\right)'
\]
with
\[
\mathbf{U}_{i1}(\mathcal{D}_i;\boldsymbol{\phi}_1)
=
\left(
\mathbf{U}_{i11}'(\mathcal{D}_i;\boldsymbol{\phi}_1),
\mathbf{U}_{i12}'(\mathcal{D}_i;\boldsymbol{\alpha}_1)
\right)'
\]
and
\[
\begin{aligned}
\mathbf{U}_{i2}\left(\mathcal{D}_i ;\boldsymbol{\gamma}_1,\boldsymbol{\phi}_2\right)
&=
\Bigl(
\mathbf{U}_{i21}^{\prime}\left(\mathcal{D}_i ; \boldsymbol{\gamma}_1, \boldsymbol{\phi}_2\right),
\mathbf{U}_{i22}^{\prime}\left(\mathcal{D}_i ; \boldsymbol{\alpha}_2\right),\\
&\qquad
\mathbf{U}_{i23}^{\prime}\left(\mathcal{D}_i ; \boldsymbol{\gamma}_1, \boldsymbol{\zeta}\right)
\Bigr)^{\prime}.
\end{aligned}
\]
Solving \eqref{equa-13} yields the estimators of the Stage~I and Stage~II causal effects. In practice, estimation proceeds sequentially: we first solve the Stage~I equations to obtain $\hat{\boldsymbol{\phi}}_{1}$ and hence $\hat{\boldsymbol{\gamma}}_{11}$, and then substitute $\hat{\boldsymbol{\gamma}}_{11}$ into the Stage~II equations to obtain $\hat{\boldsymbol{\phi}}_{2}$, including $\hat{\boldsymbol{\gamma}}_{21}$.

Let $\boldsymbol{\Omega}=(\boldsymbol{\phi}_1',\boldsymbol{\phi}_2')'$ denote the full parameter vector and $\hat{\boldsymbol{\Omega}}=(\hat{\boldsymbol{\phi}}_1',\hat{\boldsymbol{\phi}}_2')'$ the estimator obtained from \eqref{equa-13}. Under Assumptions~\ref{ass:sutva_multi}--\ref{ass:positivity_multi}, if the Stage~I model $S_1(\mathbf{Z}_{i1};\boldsymbol{\alpha}_1)$ is correctly specified, then $\hat{\boldsymbol{\gamma}}_{11}$ is consistent for the causal dose-response effect $\boldsymbol{\theta}_{1}$. Moreover, if $\hat{\boldsymbol{\gamma}}_{11}$ is consistent and either a) the Stage~II propensity score model $S_2(\mathbf{Z}_{i2};\boldsymbol{\alpha}_2)$ or b) the Stage~II imputation model $\mathbf{m}_a(\boldsymbol{\zeta}_a)$ is correct (or both), then $\hat{\boldsymbol{\gamma}}_{21}$ is consistent for the causal effect $\boldsymbol{\theta}_{2}$; see Section 1 of \textit{Supplementary Material} for related derivations.

Under Assumptions~\ref{ass:sutva_multi}--\ref{ass:positivity_multi} and standard regularity conditions \citep{van2000asymptotic}, the joint estimator $\hat{\boldsymbol{\Omega}}=(\hat{\boldsymbol{\phi}}_1',\hat{\boldsymbol{\phi}}_2')'$ is consistent and asymptotically normal. Specifically,
$
\sqrt{n}(\hat{\boldsymbol{\Omega}}-\boldsymbol{\Omega})
\overset{d}{\rightarrow}
N\bigl(0,\boldsymbol{\Sigma}(\boldsymbol{\Omega})\bigr),$
where
$
\boldsymbol{\Sigma}(\boldsymbol{\Omega})
=
\mathcal{A}^{-1}(\boldsymbol{\Omega})\,
\mathcal{B}(\boldsymbol{\Omega})\,
\{\mathcal{A}^{-1}(\boldsymbol{\Omega})\}'$ 
with $\mathcal{A}(\boldsymbol{\Omega})
=
E\left\{
-{\partial \mathbf{U}_i(\boldsymbol{\Omega})}/{\partial \boldsymbol{\Omega}'}
\right\}$ and $
\mathcal{B}(\boldsymbol{\Omega})
=
E\left\{
\mathbf{U}_i(\boldsymbol{\Omega})\mathbf{U}_i(\boldsymbol{\Omega})'
\right\}.$

Inference is based on the empirical sandwich estimator obtained by replacing $\mathcal{A}(\cdot)$ and $\mathcal{B}(\cdot)$ with their sample analogues. The resulting covariance estimator $\hat{\boldsymbol{\Sigma}}(\hat{\boldsymbol{\Omega}})$ yields standard errors, Wald confidence intervals, and Wald tests. Full derivations and proofs are given in Section 1 of the \textit{Supplementary Material}.

\subsection{Homogeneity Testing of Causal Effects}
\label{sec-3.4}

Pooling information across outcomes can sharpen inference, but such synthesis is meaningful only when the relevant causal effects are homogeneous across outcomes. We therefore develop three Wald-type tests within the proposed two-stage framework: a test for homogeneity of the causal dose-response effects among exposed individuals, a test for homogeneity of the causal effects of exposure status at the reference dose relative to no exposure, and a global test for whether both components may be regarded as common across outcomes.

\subsubsection{Homogeneity of Causal Dose-response Effects among Exposed}
To assess whether the causal dose-response effects are common across the $J$ outcomes, we consider the null hypothesis
$
H_0^{\mathrm{I}}:\gamma_{111}=\gamma_{112}=\cdots=\gamma_{11J}.
$
Let $\boldsymbol{\gamma}_{11}=(\gamma_{111},\ldots,\gamma_{11J})'$ and define
$
\boldsymbol{\delta}_{11}=\mathbf{R}\boldsymbol{\gamma}_{11},
$
where $\mathbf{R}$ is a $(J-1)\times J$ contrast matrix comparing each component of $\boldsymbol{\gamma}_{11}$ with the first. Then $H_0^{\mathrm{I}}$ is equivalent to $\boldsymbol{\delta}_{11}=\mathbf{0}$. By asymptotic normality of $\hat{\boldsymbol{\gamma}}_{11}$, the estimator $\hat{\boldsymbol{\delta}}_{11}=\mathbf{R}\hat{\boldsymbol{\gamma}}_{11}$ is asymptotically multivariate normal with covariance matrix estimated by
$
\widehat{\boldsymbol{\Sigma}}_{\boldsymbol{\delta}_{11}}
=
\mathbf{R}\,\widehat{\mathrm{cov}}(\hat{\boldsymbol{\gamma}}_{11})\,\mathbf{R}'.
$
The Wald statistic
$
\Lambda_{\mathrm{I}}
=
\hat{\boldsymbol{\delta}}_{11}'
\widehat{\boldsymbol{\Sigma}}_{\boldsymbol{\delta}_{11}}^{-1}
\hat{\boldsymbol{\delta}}_{11}
$
has an asymptotic chi-squared distribution with $(J-1)$ degrees of freedom under $H_0^{\mathrm{I}}$.
\subsubsection{Homogeneity of Causal Effects of Exposure Status}
Homogeneity of the exposure-status effects, $H_0^{\mathrm{II}}:\gamma_{211}=\gamma_{212}=\cdots=\gamma_{21J}$, is assessed analogously. With $\boldsymbol{\gamma}_{21}=(\gamma_{211},\ldots,\gamma_{21J})'$, $\boldsymbol{\delta}_{21}=\mathbf{R}\boldsymbol{\gamma}_{21}$, and $\widehat{\boldsymbol{\Sigma}}_{\boldsymbol{\delta}_{21}}=\mathbf{R}\,\widehat{\mathrm{cov}}(\hat{\boldsymbol{\gamma}}_{21})\,\mathbf{R}'$, the Wald statistic $\Lambda_{\mathrm{II}}=\hat{\boldsymbol{\delta}}_{21}'\widehat{\boldsymbol{\Sigma}}_{\boldsymbol{\delta}_{21}}^{-1}\hat{\boldsymbol{\delta}}_{21}$ has an asymptotic chi-squared distribution with $(J-1)$ degrees of freedom under $H_0^{\mathrm{II}}$.

\subsubsection{Global Homogeneity Test}
The global null of simultaneous homogeneity is $H_0^{\mathrm{G}}:\boldsymbol{\delta}_{11}=\mathbf{0}$ and $\boldsymbol{\delta}_{21}=\mathbf{0}$. Let $\boldsymbol{\delta}=(\boldsymbol{\delta}_{11}',\boldsymbol{\delta}_{21}')'$ and let $\mathbf{R}_{\mathrm{G}}$ be the block-diagonal matrix with $\mathbf{R}$ in each diagonal block, so $\boldsymbol{\delta}=\mathbf{R}_{\mathrm{G}}(\boldsymbol{\gamma}_{11}',\boldsymbol{\gamma}_{21}')'$ with estimated covariance $\widehat{\boldsymbol{\Sigma}}_{\boldsymbol{\delta}}=\mathbf{R}_{\mathrm{G}}\,\widehat{\mathrm{cov}}\bigl\{(\hat{\boldsymbol{\gamma}}_{11}',\hat{\boldsymbol{\gamma}}_{21}')'\bigr\}\mathbf{R}_{\mathrm{G}}'$. The global Wald statistic $\Lambda_{\mathrm{G}}=\hat{\boldsymbol{\delta}}'\widehat{\boldsymbol{\Sigma}}_{\boldsymbol{\delta}}^{-1}\hat{\boldsymbol{\delta}}$ has a chi-squared distribution with $2(J-1)$ degrees of freedom under $H_0^{\mathrm{G}}$.

When homogeneity is not rejected, pooled summaries of the corresponding causal effects may be reported to provide an overall impact of exposure across outcomes. The finite sample performance of the proposed two-stage estimators and within-cohort homogeneity tests is examined in Section 2 of the \textit{Supplementary Material}.

\section{Evidence Synthesis with Multiple Cohort Studies}
\label{sec-4}
To this point, we have focused on the analysis of a single cohort, but interest lies in synthesizing evidence across studies when they are available. This is particularly appealing when small effects with important public health implications are difficult to detect. Here we define cohort-specific estimators of causal effects for each domain and develop homogeneity tests across cohorts within domain and across domains to support an overall causal effect.

\subsection{Notation}
\label{sec-4.1:multicohort-notation}

Suppose data are available from $S$ independent cohorts. Let $\mathbf{Y}_{si}=(Y_{si1},\dots,Y_{siJ})'$ denote the $J$-dimensional vector of outcomes for individual $i$ in cohort $s$, where $i=1,\ldots,n_s$ and $s=1,\ldots,S$. Let $A_{si}$ denote exposure status, $T_{si}$ denote the non-negative exposure level when $A_{si}=1$, and $D_{si}=\log T_{si}$ for exposed individuals.

Then let $
\hat{\boldsymbol{\gamma}}_s
=
(\hat{\boldsymbol{\gamma}}_{11,s}',\hat{\boldsymbol{\gamma}}_{21,s}')'.
$ with 
$
\hat{\boldsymbol{\gamma}}_{11,s}
=
(\hat{\gamma}_{11,s1},\dots,\hat{\gamma}_{11,sJ})'$
and
$
\hat{\boldsymbol{\gamma}}_{21,s}
=
(\hat{\gamma}_{21,s1},\dots,\hat{\gamma}_{21,sJ})'
$. Let $\hat{\gamma}_{11,sj}$ estimate the causal dose-response effect among exposed individuals for domain $j$ in cohort $s$, and $\hat{\gamma}_{21,sj}$ estimates the causal effect comparing exposure at the reference dose $c$ with no exposure for domain $j$ in cohort $s$, obtained from the two-stage procedure in Section~\ref{sec-3.3}, $j=1,\ldots,J$, and let
$\hat{\boldsymbol{\Sigma}}_s=\widehat{\mathrm{cov}}
\bigl(\hat{\boldsymbol{\gamma}}_s\bigr)$
denote the estimated covariance matrix for the cohort-specific effect estimators.

To assess homogeneity across cohorts within a fixed domain $j$, it is convenient to collect the cohort-specific estimators of 
$\boldsymbol{\gamma}_{11,j}
=
(\gamma_{11,1j},\dots,\gamma_{11,Sj})'$
and
$\boldsymbol{\gamma}_{21,j}=(\gamma_{21,1j},\dots,\gamma_{21,Sj})'$ in 
$\hat{\boldsymbol{\gamma}}_{11,j}=(\hat{\gamma}_{11,1j},\dots,\hat{\gamma}_{11,Sj})'$
and
$ \hat{\boldsymbol{\gamma}}_{21,j}=(\hat{\gamma}_{21,1j},\dots,\hat{\gamma}_{21,Sj})'.$
If effects appear homogeneous across cohorts within a domain, we obtain the pooled cohort-specific estimators $\tilde{\gamma}_{11,j}$ and $\tilde{\gamma}_{21,j}$ for Stage~I and Stage~II, respectively. These pooled domain-specific estimators can then be assembled across domains as $\tilde{\boldsymbol{\gamma}}_{11}=(\tilde{\gamma}_{11,1},\dots,\tilde{\gamma}_{11,J})'$ and
$
\tilde{\boldsymbol{\gamma}}_{21}
=
(\tilde{\gamma}_{21,1},\dots,\tilde{\gamma}_{21,J})',
$
which provide the basis for a second layer of homogeneity assessment across domains. 

We use hats for cohort-specific estimators, tildes for pooled estimators within domain, and bars for overall pooled domain-specific estimators. Section~\ref{sec-4.2:cohort-homogeneity} develops tests of homogeneity across cohorts within domain, and Section~\ref{sec-4.3:domain-homogeneity} considers homogeneity across domains based on the pooled domain-specific estimators.

\subsection{Assessing Homogeneity across Cohorts within a Domain}
\label{sec-4.2:cohort-homogeneity}

Applying the two-stage procedure of Section~\ref{sec-3} separately within each cohort yields cohort-specific Stage~I and Stage~II estimators; we then assess, for each domain, whether these are similar enough to support pooling across cohorts.

Let
$
\hat{\boldsymbol{\Sigma}}_{11,j}
=
\widehat{\mathrm{cov}}(\hat{\boldsymbol{\gamma}}_{11,j})
$
and
$
\hat{\boldsymbol{\Sigma}}_{21,j}
=
\widehat{\mathrm{cov}}(\hat{\boldsymbol{\gamma}}_{21,j})
$
denote their estimated covariance matrices, $j=1,\ldots,J$. To assess homogeneity of the Stage~I effects across cohorts for domain $j$, consider
$
H^{(C)}_{01j}:\gamma_{11,1j}=\gamma_{11,2j}=\cdots=\gamma_{11,Sj}.
$
Let $\mathbf{C}_S$ be an $(S-1)\times S$ contrast matrix satisfying
$\mathbf{C}_S\mathbf{1}_S=\mathbf{0}$. Then $H^{(C)}_{01j}$ is equivalent to $\mathbf{C}_S\boldsymbol{\gamma}_{11,j}=\mathbf{0}$, and we use the Wald statistic
$
Q_{11,j}=
(\mathbf{C}_S\hat{\boldsymbol{\gamma}}_{11,j})'
\left\{
\mathbf{C}_S\hat{\boldsymbol{\Sigma}}_{11,j}\mathbf{C}_S'
\right\}^{-1}
(\mathbf{C}_S\hat{\boldsymbol{\gamma}}_{11,j}).
$
Under $H^{(C)}_{01j}$ and standard regularity conditions \citep{doob1935limiting}, the Wald statistic $Q_{11,j}$ is asymptotically chi-squared with $(S-1)$ degrees of freedom. To assess the homogeneity of the Stage~II effects across cohorts for domain $j$, we test 
$
H^{(C)}_{02j}:\gamma_{21,1j}=\gamma_{21,2j}=\cdots=\gamma_{21,Sj},
$
using the Wald statistic
$
Q_{21,j}
=
(\mathbf{C}_S\hat{\boldsymbol{\gamma}}_{21,j})'
\left\{
\mathbf{C}_S\hat{\boldsymbol{\Sigma}}_{21,j}\mathbf{C}_S'
\right\}^{-1}
(\mathbf{C}_S\hat{\boldsymbol{\gamma}}_{21,j}),
$
which is asymptotically distributed as $\chi^2_{S-1}$ under $H^{(C)}_{02j}$.

As the semi-continuous exposure gives rise to two causal estimands, it is also useful to assess whether the pair of cohort-specific effects is jointly homogeneous within domain $j$. To this end, we let
$
\boldsymbol{\eta}_j
=
(\boldsymbol{\gamma}_{11,j}',\boldsymbol{\gamma}_{21,j}')'
$
and
$
\hat{\boldsymbol{\eta}}_j
=
(\hat{\boldsymbol{\gamma}}_{11,j}',\hat{\boldsymbol{\gamma}}_{21,j}')'
$
denote the vectors of true and estimated cohort-specific effects from both stages, with corresponding covariance matrix
$
\hat{\boldsymbol{\Sigma}}_{j}
=
\widehat{\mathrm{cov}}(\hat{\boldsymbol{\eta}}_j).
$
We test the global null hypothesis
$
H^{(C)}_{0j}:
\gamma_{11,1j}=\cdots=\gamma_{11,Sj}$ and $
\gamma_{21,1j}=\cdots=\gamma_{21,Sj}.
$ Let $\mathbf{A}_S$ denote the block-diagonal matrix with two copies of $\mathbf{C}_S$ on the diagonal. Then $H^{(C)}_{0j}$ is equivalent to $\mathbf{A}_S\boldsymbol{\eta}_j=\mathbf{0}$, and the corresponding Wald statistic $Q_j$ is
$
(\mathbf{A}_S\hat{\boldsymbol{\eta}}_j)'
\left\{
\mathbf{A}_S\hat{\boldsymbol{\Sigma}}_{j}\mathbf{A}_S'
\right\}^{-1}
(\mathbf{A}_S\hat{\boldsymbol{\eta}}_j),
$
which is asymptotically distributed as $\chi^2_{2(S-1)}$ under $H^{(C)}_{0j}$.

When the evidence against homogeneity is weak, a common domain-specific causal dose-response effect can be obtained by inverse-variance weighting
$$
\tilde{\gamma}_{11,j}
=
(\mathbf{1}_S'\hat{\boldsymbol{\Sigma}}_{11,j}^{-1}\mathbf{1}_S)^{-1}
\mathbf{1}_S'\hat{\boldsymbol{\Sigma}}_{11,j}^{-1}\hat{\boldsymbol{\gamma}}_{11,j},
$$
with 
$
\widehat{\mathrm{var}}(\tilde{\gamma}_{11,j})
=
(\mathbf{1}_S'\hat{\boldsymbol{\Sigma}}_{11,j}^{-1}\mathbf{1}_S)^{-1}
$ \citep{cochran1954combination,hartung2011statistical}. Under the common-effect model, this is the most precise linear unbiased estimator \citep{hedges2014statistical}. Analogously, when $H^{(C)}_{02j}$ is not rejected, the pooled Stage~II effect is
$
\tilde{\gamma}_{21,j}
=
(\mathbf{1}_S'\hat{\boldsymbol{\Sigma}}_{21,j}^{-1}\mathbf{1}_S)^{-1}
\mathbf{1}_S'\hat{\boldsymbol{\Sigma}}_{21,j}^{-1}\hat{\boldsymbol{\gamma}}_{21,j},
$
with $
\widehat{\mathrm{var}}(\tilde{\gamma}_{21,j})
=
(\mathbf{1}_S'\hat{\boldsymbol{\Sigma}}_{21,j}^{-1}\mathbf{1}_S)^{-1}.
$
The resulting pooled estimates form the basis for the second layer of evidence synthesis across domains in Section~\ref{sec-4.3:domain-homogeneity}.

\subsection{Homogeneity across domains and overall pooled inference}
\label{sec-4.3:domain-homogeneity}
We now assess whether the pooled domain-specific effects are sufficiently similar across domains to support an overall summary. Because the evidence for within-domain homogeneity may differ between Stage~I and Stage~II, the set of domains eligible for cross-domain pooling need not be the same for the two causal components.

Let $\mathcal{J}_1 \subseteq \{1,\dots,J\}$ denote the set of domains for which dose-response effects are synthesized with $K_1 = |\mathcal{J}_1|$. Similarly, let $\mathcal{J}_2 \subseteq \{1,\dots,J\}$ denote the set of domains for which the exposure-status effects are pooled, and let $K_2 = |\mathcal{J}_2|$. We then let 
$
\tilde{\boldsymbol{\gamma}}_{11}
=
(\tilde{\gamma}_{11,j}: j \in \mathcal{J}_1)'
$
and $\hat{\boldsymbol{\Omega}}_{11} = \widehat{\mathrm{cov}}(\tilde{\boldsymbol{\gamma}}_{11})$ denote its estimated covariance matrix, and 
$
\tilde{\boldsymbol{\gamma}}_{21}
=
(\tilde{\gamma}_{21,j}: j \in \mathcal{J}_2)' ,
$
and 
$
\hat{\boldsymbol{\Omega}}_{21} = \widehat{\mathrm{cov}}(\tilde{\boldsymbol{\gamma}}_{21}) .
$

To assess homogeneity of the pooled Stage~I effects across domains, we consider
$
H^{(D)}_{01}: \tilde{\gamma}_{11,j_1} = \tilde{\gamma}_{11,j_2}$ for all $j_1,j_2 \in \mathcal{J}_1$. 
Let $\mathbf{R}_{K_1}$ be a $(K_1-1)\times K_1$ contrast matrix satisfying
$\mathbf{R}_{K_1}\mathbf{1}_{K_1} = \mathbf{0}$. Then $H^{(D)}_{01}$ is equivalent to
$\mathbf{R}_{K_1}\tilde{\boldsymbol{\gamma}}_{11} = \mathbf{0}$, and we use the Wald statistic
$
Q^{(D)}_{11}
=
(\mathbf{R}_{K_1}\tilde{\boldsymbol{\gamma}}_{11})'
\left\{
\mathbf{R}_{K_1}\hat{\boldsymbol{\Omega}}_{11}\mathbf{R}_{K_1}'
\right\}^{-1}
(\mathbf{R}_{K_1}\tilde{\boldsymbol{\gamma}}_{11}) .
$
Under $H^{(D)}_{01}$ and standard regularity conditions, $Q^{(D)}_{11}$ has an asymptotic chi-squared distribution with $(K_1-1)$ degrees of freedom.

Similarly, to assess homogeneity of the pooled Stage~II effects across domains, we test
$
H^{(D)}_{02}: \tilde{\gamma}_{21,j_1} = \tilde{\gamma}_{21,j_2}$
for all $j_1,j_2 \in \mathcal{J}_2$
using the Wald statistic
$
Q^{(D)}_{21}
=
(\mathbf{R}_{K_2}\tilde{\boldsymbol{\gamma}}_{21})'
\left\{
\mathbf{R}_{K_2}\hat{\boldsymbol{\Omega}}_{21}\mathbf{R}_{K_2}'
\right\}^{-1}
(\mathbf{R}_{K_2}\tilde{\boldsymbol{\gamma}}_{21}) ,
$
which has an asymptotic chi-squared distribution with $(K_2-1)$ degrees of freedom under $H^{(D)}_{02}$.

When $K_1<2$ or $K_2<2$, the corresponding cross-domain homogeneity test is not performed because fewer than two pooled domain-specific effects are available. When the evidence against cross-domain homogeneity is weak, we summarize the overall Stage~I effect by inverse-variance weighting of the pooled domain-specific estimators:
$
\bar{\gamma}_{11}
=
(\mathbf{1}_{K_1}'\hat{\boldsymbol{\Omega}}_{11}^{-1}\mathbf{1}_{K_1})^{-1}
\mathbf{1}_{K_1}'\hat{\boldsymbol{\Omega}}_{11}^{-1}\tilde{\boldsymbol{\gamma}}_{11},
$
with estimated variance
$
\hat{\mathrm{var}}(\bar{\gamma}_{11})
=
(\mathbf{1}_{K_1}'\hat{\boldsymbol{\Omega}}_{11}^{-1}\mathbf{1}_{K_1})^{-1}.
$
Analogously, when $H^{(D)}_{02}$ is not rejected, the overall pooled Stage~II effect is
$
\bar{\gamma}_{21}
=
(\mathbf{1}_{K_2}'\hat{\boldsymbol{\Omega}}_{21}^{-1}\mathbf{1}_{K_2})^{-1}
\mathbf{1}_{K_2}'\hat{\boldsymbol{\Omega}}_{21}^{-1}\tilde{\boldsymbol{\gamma}}_{21},
$
with estimated variance
$
\hat{\mathrm{var}}(\bar{\gamma}_{21})
=
(\mathbf{1}_{K_2}'\hat{\boldsymbol{\Omega}}_{21}^{-1}\mathbf{1}_{K_2})^{-1}.
$

Thus, evidence synthesis proceeds hierarchically: we assess homogeneity across cohorts within domain, pool when appropriate, and then assess homogeneity across the pooled domain-specific effects. In Section 2 of the \textit{Supplementary Material} we report on the findings from extensive simulation studies examining the finite sample performance of the estimators developed here.

\section{The Effects of PAE on Childhood Cognition}
\label{sec-5}
We apply the proposed methods to study the effects of prenatal alcohol exposure on childhood cognition using the harmonized data described in Section~\ref{sec-2}. The analysis focuses on three domain-level outcomes: executive function (EF), mathematics achievement (Math), and learning and memory (LM). Our primary goal is to examine whether the Stage I and Stage II effects show similar patterns across domains and cohorts. As we show below, the application suggests that Stage~I effects are relatively stable across domains and cohorts, whereas Stage~II effects are more variable across cohorts, particularly for EF. We proceed in two steps. First, we illustrate the single-cohort framework of Section~\ref{sec-3} using the Detroit cohort. We then apply the multi-cohort synthesis framework of Section~\ref{sec-4} to combine evidence across the five cohorts.

\subsection{Analysis of a Single-cohort Framework}
\label{sec-5.1}

We first illustrate the single-cohort framework of Section~\ref{sec-3} using the Detroit cohort. This cohort provides a natural starting point because it extends the application in \cite{wang2025two} to the multiple-outcome setting considered here. As in Section~\ref{sec-3}, we target two causal effects within this cohort: the Stage~I causal dose-response effect among exposed mothers and the Stage~II causal effect comparing prenatal alcohol exposure at the reference dose of 1 oz of absolute alcohol per day with no exposure. We use 1 oz of absolute alcohol per day as the reference dose to provide a clinically interpretable contrast with no exposure and a common target across cohorts. The Detroit analysis uses the cohort-specific baseline covariates available for confounding adjustment, and missing baseline covariates and outcomes are handled using multiple imputation as described in Section 4 of Supplementary Material. Additional details on the Detroit propensity score specifications and diagnostic summaries are provided in Section 5 of \textit{Supplementary Material}.

Figure~\ref{fig-detroit} displays the Detroit-specific domain-level estimates and corresponding pooled effects across the three cognitive domains. Panel A shows the Stage I estimates, corresponding to the causal dose-response effects among exposed mothers. The estimated effects are negative for all three domains, with estimates of $-0.731$ for EF, $-0.243$ for Math, and $-0.399$ for LM. These results suggest that, within the Detroit cohort, higher alcohol intake among exposed mothers leads to lower cognitive scores across all three domains, although the estimated effects are modest in magnitude and the confidence intervals are relatively wide. Panel B shows the Stage II estimates, corresponding to the effect of PAE at 1 oz of AA per day relative to no exposure. These estimated effects are also negative across all domains, with estimates of $-8.248$ for EF, $-4.751$ for Math, and $-5.940$ for LM, indicating an adverse effect of PAE on childhood cognition in Detroit.

\begin{figure}[t]
\centering
\includegraphics[width=0.8\textwidth]{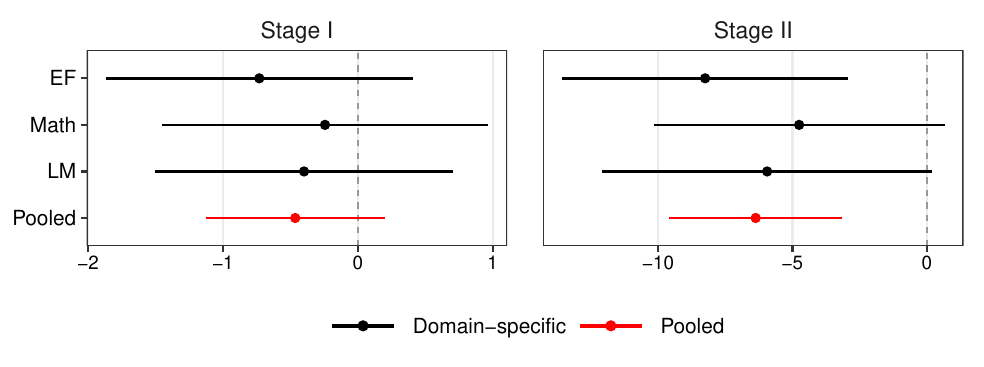}
\caption{Estimated domain-specific and pooled PAE effects with 95\% confidence intervals in the Detroit cohort.}
\label{fig-detroit}
\end{figure}
We next assess whether the estimated effects are homogeneous across EF, Math, and LM to support pooled inference within the Detroit cohort, following the homogeneity tests developed in Section~\ref{sec-3.4}. For the Stage I causal dose-response effects among exposed, the homogeneity test yields $Q=0.947$ on $2$ degrees of freedom with $p$-value of $0.623$, providing little evidence against a common effect across the three domains. For the Stage II effects, the corresponding test gives $Q=2.066$ on $2$ degrees of freedom with $p$-value of $0.356$, again providing little evidence against homogeneity across outcomes. A global test assessing homogeneity across domains jointly for the Stage I and Stage II effects gives $Q=2.291$ on $4$ degrees of freedom with $p$-value of $0.682$, providing little evidence against the overall homogeneity assumption. These results support pooling across domains for both causal effects.

Using inverse-variance weighting \citep{cochran1954combination,hartung2011statistical} across the three domain-specific estimates, the pooled Stage I effect in Detroit is $-0.464$ with standard error $0.338$ and 95\% confidence interval $(-1.127,\,0.199)$. The pooled Stage II effect is $-6.367$ with standard error $1.646$ and 95\% confidence interval $(-9.593,\,-3.140)$. This indicates that prenatal alcohol exposure at 1 oz of absolute alcohol per day, compared with no exposure, causes a lower average cognitive score across the three domains in the Detroit cohort. Both Stage~I and Stage~II estimates are negative across EF, Math, and LM, and the corresponding homogeneity tests provide little evidence against pooling within cohort. At the same time, the Stage~II effects are larger in magnitude than the Stage~I effects, suggesting that the effect of drinking at 1 oz AA per day compared to unexposed may be more pronounced than the incremental dose-response effect among exposed mothers. This single-cohort analysis motivates the broader multi-cohort investigation, where we examine whether the relative stability of Stage~I effects and the potentially greater variability of Stage~II effects persist across studies.

\subsection{Implementation of the Multi-cohort Framework}
\label{sec-5.2}
We next apply the multi-cohort framework of Section~\ref{sec-4} to synthesize evidence across the five cohorts: Detroit, Seattle, Atlanta, and Pittsburgh Cohorts 1 and 2. A central question is whether the relatively homogeneous pattern seen in Detroit extends across cohorts, and in particular, whether the Stage I effects are more stable than the Stage II effects. Within each cohort, we use the same two-stage procedure described in Section~\ref{sec-3} and illustrated for Detroit in Section~\ref{sec-5.1}. In particular, we estimate the Stage I and Stage II effects for each of the three domains (EF, Math, LM). To accommodate differences in covariate availability across cohorts, the generalized propensity score model for dose among exposed mothers and the propensity score model for exposure status are fit separately within each cohort using the cohort-specific set of baseline confounders. The cohort-specific propensity score specifications, together with diagnostic summaries for the Stage I and Stage II models, are provided in Section 5 of \textit{Supplementary Material}. Missing baseline covariates and cognitive outcomes are handled within each cohort using multiple imputation by chained equations with 20 imputations, and the resulting estimates and covariance matrices are combined using Rubin's rules \citep{rubin1976inference,little2019statistical}. Additional details on the cohort-specific imputation procedure are provided in \textit{Supplementary Material}. In the notation of Section~\ref{sec-4}, the cohort-specific estimates correspond to the hat quantities, the pooled domain-specific estimates across cohorts correspond to the tilde quantities, and the overall pooled estimates across domains correspond to the bar quantities; in this section, we describe these primarily in words.

\begin{figure}[t]
\centering
\includegraphics[width=1\textwidth]{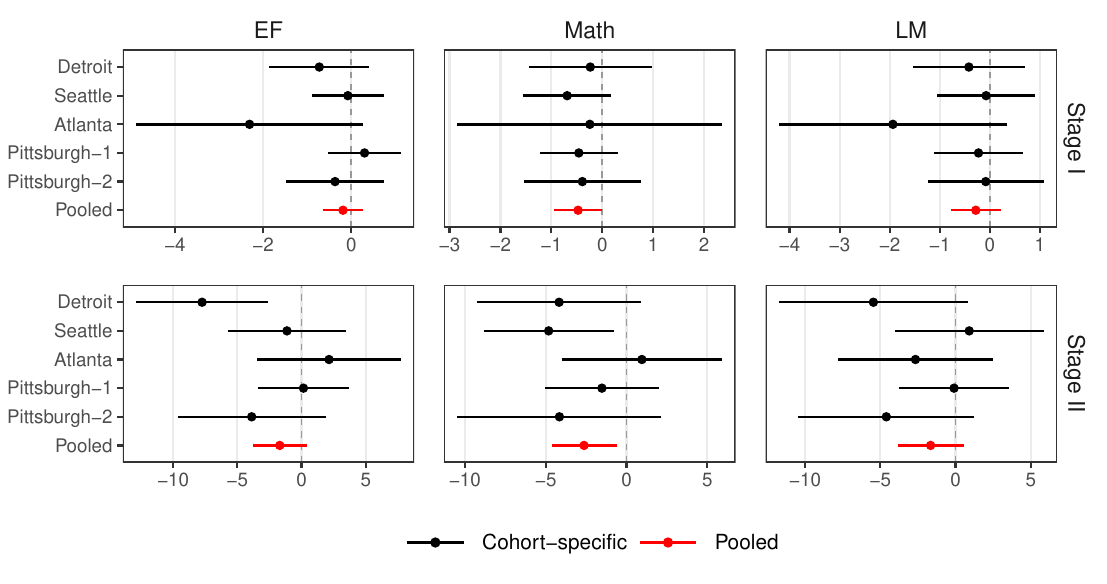}
\caption{Estimated PAE effects by cohort and cognitive domain with 95\% confidence intervals.}
\label{forestplot}
\end{figure}

Figure~\ref{forestplot} displays the point estimates and the 95\% confidence intervals of the cohort-specific two-stage estimates of the effects of PAE on EF, Math, and LM; see Table 11 in the \textit{Supplementary Material} for numerical results. For Stage~I, the estimated causal dose-response effects among exposed mothers are generally negative across cohorts and domains, indicating worse cognitive performance with increasing alcohol consumption among mothers who drink during pregnancy. The estimated Stage~I effects vary in magnitude across cohorts, with the strongest negative estimates appearing in Atlanta for EF and LM, while the estimates in the remaining cohorts are smaller in magnitude and often close to zero. For Stage~II, the estimated effects comparing exposure at 1 oz of absolute alcohol per day with no exposure are also predominantly negative, but they show slightly greater between-cohort variation than the Stage~I estimates. Detroit and Pittsburgh Cohort~2 exhibit consistently negative Stage~II estimates across EF, Math, and LM, while estimates in Pittsburgh Cohort~1 are close to zero and those in Atlanta are centered near zero with confidence intervals overlapping zero for all three domains. Overall, the cohort-specific results suggest that the causal dose-response effect among exposed mothers is more stable across cohorts than the causal effect of drinking 1 oz AA/day compared with no exposure.  
 
We next assess whether the cohort-specific effects are sufficiently similar to support evidence synthesis across cohorts within each domain, following the homogeneity tests developed in Section~\ref{sec-4}. For each domain, we test homogeneity of the Stage I effects, homogeneity of the Stage II effects, and a within-domain global null.

Table~\ref{tab:homogeneity_tests_5cohorts_new} reports the resulting $p$-values together with pooled estimates when homogeneity is not rejected. For Stage~I, there is little evidence against homogeneity across cohorts for any of the three domains, with $p$-values of $0.288$, $0.980$, and $0.668$ for EF, Math, and LM, respectively. Accordingly, we report pooled Stage~I estimates within each domain: $-0.188$ for EF, $-0.475$ for Math, and $-0.276$ for LM with similar standard errors of around $0.24$. These pooled Stage~I estimates are all negative, suggesting that, among exposed mothers, higher alcohol consumption causes lower cognitive scores across the three domains, especially for Math. For Stage~II, there is also little evidence against homogeneity across cohorts for any of the three domains, with $p$-values of $0.055$, $0.364$, and $0.319$ for EF, Math, and LM, respectively, although the evidence for EF is closer to the conventional significance threshold. Accordingly, we report pooled Stage~II estimates within each domain: $-1.710$ for EF, $-2.680$ for Math, and $-1.736$ for LM, with corresponding standard errors of $1.081$, $1.036$, and $1.117$, respectively.
\begin{table}[t]
\scriptsize
\centering
\caption{Within-domain homogeneity tests across five cohorts and corresponding pooled estimates.}
\label{tab:homogeneity_tests_5cohorts_new}
\begin{tabular}{llcccc}
\hline
Domain & Test & df & $p$-value & Pooled estimate & SE \\
\hline
\multirow[t]{3}{*}{Executive function}
& $H^{(C)}_{011}$ & 4 & 0.288 & $-0.188$ & 0.235 \\
& $H^{(C)}_{021}$ & 4 & 0.055 & $-1.710$ & 1.081 \\
& $H^{(C)}_{01}$  & 8 & 0.106 & -- & -- \\
\hline
\multirow[t]{3}{*}{Math}
& $H^{(C)}_{012}$ & 4 & 0.980 & $-0.475$ & 0.237 \\
& $H^{(C)}_{022}$ & 4 & 0.364 & $-2.680$ & 1.036 \\
& $H^{(C)}_{02}$  & 8 & 0.716 & -- & -- \\
\hline
\multirow[t]{3}{*}{Learning and memory}
& $H^{(C)}_{013}$ & 4 & 0.668 & $-0.276$ & 0.253 \\
& $H^{(C)}_{023}$ & 4 & 0.319 & $-1.736$ & 1.117 \\
& $H^{(C)}_{03}$  & 8 & 0.407 & -- & -- \\
\hline
\end{tabular}
\end{table}

We then apply the procedure of evidence synthesis in Section~\ref{sec-4.3:domain-homogeneity} to the pooled domain-specific effects. For Stage~I, the pooled effects for EF, Math, and LM are sufficiently similar across domains, with a homogeneity test $p$-value of $0.354$. We therefore summarize the Stage~I results by pooling the three domain-specific Stage~I effects, yielding an overall estimate of $-0.304$ with standard error $0.199$. These pooled Stage~I estimates are all negative, suggesting that, among exposed mothers, higher alcohol consumption may cause lower cognitive scores across all three domains, with the largest estimated effect observed for Math.

\begin{figure}[t] 
\centering
\includegraphics[width=0.865\textwidth]{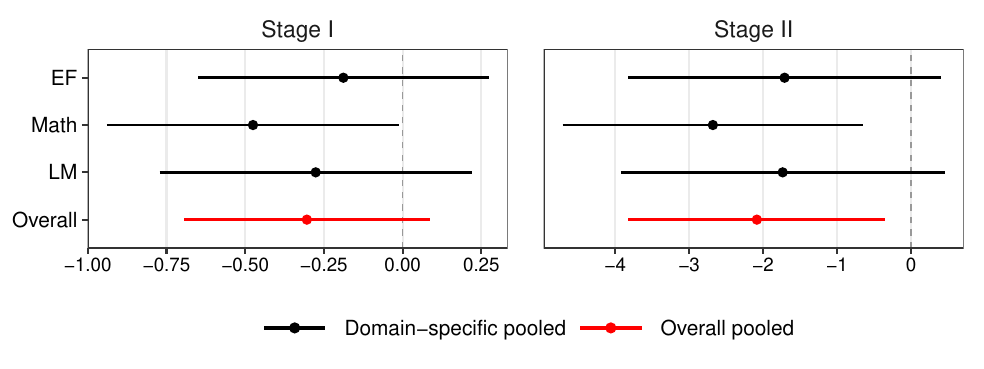}
\caption{Estimated pooled PAE effects by cognitive domain across cohorts with 95\% confidence intervals.}
\label{overall-forestplot}
\end{figure}

For Stage~II, the cohort-level homogeneity results support pooling across cohorts for EF, Math, and LM. We therefore assess homogeneity across the pooled effects from all three domains. The resulting test gives $p=0.523$, providing little evidence against homogeneity across domains. Pooling the EF, Math, and LM Stage~II estimates yields an overall effect of $-2.085$ with standard error $0.887$. This estimate indicates that prenatal alcohol exposure at 1 oz of absolute alcohol per day is associated with a lower average cognitive score across EF, Math, and LM relative to no exposure. Figure~\ref{overall-forestplot} summarizes the pooled domain-specific estimates across cohorts, together with the corresponding overall pooled estimates across domains, for both Stage~I and Stage~II.

Taken together, these findings show that the multi-cohort analysis broadly reinforces the pattern seen in Detroit while also revealing important between-cohort differences. Across cohorts, the Stage~I dose-response effects among exposed mothers are consistently negative and sufficiently homogeneous to support pooling both within domain and across domains. In contrast, the Stage II effects show greater heterogeneity, particularly for EF, although the pooled effects for EF, Math, and LM are all negative. Substantively, these findings suggest that increasing alcohol intake among mothers who drink during pregnancy leads to broadly adverse cognitive consequences, while the contrast between drinking at 1 ounce of AA per day and no exposure may be more sensitive to cohort-specific differences in study population or exposure distribution. Figure~\ref{overall-forestplot} illustrates that the pooled Stage~I estimates are similar in magnitude across EF, Math, and LM, whereas the pooled Stage~II estimates are more variable across domains. The observed between-cohort variation in Stage~II effects may reflect differences in cohort composition, exposure distributions, measured confounders, or residual sampling variability. 

\section{Discussion}
\label{sec-7}
This article develops a two-stage framework for causal inference with semi-continuous exposures and multiple correlated continuous outcomes, and extends it to a multi-cohort setting for evidence synthesis. The framework answers two causal questions: the Stage~I dose-response effect among exposed individuals and the Stage~II effect of exposure at a fixed reference dose compared with no exposure. By combining generalized estimating equations with AIPW in Stage~II, the proposed approach yields interpretable causal effect estimates while accommodating dependence across outcomes \citep{liang1986longitudinal}. We further develop homogeneity tests to assess if these effects are sufficiently similar across outcomes within a cohort and across cohorts and domains.

The methodology is motivated by the study of PAE and childhood cognition, with interest in executive function, mathematics achievement, and learning and memory. In the application, the proposed approach decomposes the effects of PAE into the Stage I and Stage II effects. The multi-cohort extension synthesizes evidence across five cohorts while formally assessing whether pooling is appropriate. Substantively, the application suggests that prenatal alcohol exposure has broadly adverse effects on childhood cognition. Across cohorts, the Stage I effects are generally negative and relatively stable, while the Stage II effects are also negative overall but show greater between-cohort variability. Combining information across cohorts and domains, where supported by the homogeneity tests, strengthened the evidence.

Several limitations should be noted. First, estimation of the Stage~I causal dose-response effect relies on correct specification of the generalized propensity score model for the continuous exposure among exposed individuals, so misspecification at this stage may affect both Stage~I and subsequent Stage~II inference. This motivates future work on more flexible or doubly robust methods for Stage~I estimation, for example, using semiparametric or nonparametric approaches to estimate continuous treatment effects \citep{kennedy2017non,stringer2024semi}. Second, the proposed estimating equations are developed for continuous outcomes and use a working independence structure; although sandwich-based inference remains robust to misspecification of the working correlation, efficiency may be reduced when outcome dependence is substantial \citep{liang1986longitudinal}. Third, causal interpretation relies on consistency, exchangeability, and positivity and remains vulnerable to unmeasured confounding, while cross-cohort differences in measurement, design, and population may further affect the pooled results.

\section*{Acknowledgments}
The authors wish to acknowledge the principal investigators — Claire D. Coles (Atlanta Cohorts 1 and 2), Gale A. Richardson (Pittsburgh Cohort 2), the late Ann P. Streissguth and Heather Carmichael Olson (Seattle) — and their study teams for generously contributing the PAE data used in this article.
\section*{Disclosure statement}\label{disclosure-statement}

No potential conflict of interest was reported by the authors.

\section*{Funding}
This research was supported by NIH/NIAAA grant R01 AA025905 and the Lycaki-Young Fund (State of Michigan). Data collection was supported by NIH grants (AA awarded by NIAAA, DA by NIDA, HD by NICHD) as follows — Detroit: R01 AA06966, R01 AA09524, P50 AA07606, R21 DA021034; Pittsburgh: R01 AA06390, R01 AA06666, R01 AA14215, R01 AA18116, HD036890, R01 DA00090, R01 DA03209, R01 DA03874, R01 DA17786, R01 DA05460, R01 DA06839, R01 DA08916, R01 DA12401; Atlanta: R01 AA08105, R01 AA10108, R01 AA13272, R01 DA0737362, and Georgia State Contract DHR 427 93 05050762; Seattle: R01 AA001455. Additional support came from the Australian Research Council Centre of Excellence for Mathematical and Statistical Frontiers (CE140100049; L. M. Ryan), NSERC Discovery Grants (R. J. Cook, RGPIN-2017-04207; Y. Zhu, RGPIN-2017-04064), and the Canadian Institutes of Health Research (R. J. Cook and L. M. Ryan; FRN PJT-180551).

\section*{Data Availability Statement}\label{data-availability-statement}
The data are not publicly available because they contain sensitive human-subjects information, but may be available from the original cohort investigators upon reasonable request, subject to institutional and ethical approvals.

  \bibliography{bibliography.bib}

\end{document}